\documentclass[prl,twocolumn,twoside,floatfix,showpacs]{revtex4}
\usepackage{graphicx}
\usepackage{graphicx,amsfonts}
\usepackage{epsfig,amsmath}
\usepackage{verbatim,pslatex}

\begin{document}

\newcommand{\atanh}
{\operatorname{atanh}}
\newcommand{\ArcTan}
{\operatorname{ArcTan}}
\newcommand{\ArcCoth}
{\operatorname{ArcCoth}}
\newcommand{\Erf}
{\operatorname{Erf}}
\newcommand{\Erfi}
{\operatorname{Erfi}}
\newcommand{\Ei}
{\operatorname{Ei}}

\title{Strong Disorder Fixed Point in the Dissipative Random
  Transverse Field Ising Model}
\author{Gr\'egory Schehr}
\affiliation{Theoretische Physik Universit\"at des Saarlandes
66041 Saarbr\"ucken Germany}
\author{Heiko Rieger}
\affiliation{Theoretische Physik, Universit\"at des Saarlandes,
66041 Saarbr\"ucken, Germany}

\date{\today}

\begin{abstract}
The interplay between disorder, quantum fluctuations and dissipation
is studied in the random transverse Ising chain coupled to a
dissipative Ohmic bath with a real space renormalization group.  A
typically very large length scale, $L^{*}$, is identified above which
the physics of frozen clusters dominates. Below $L^*$ a strong
disorder fixed point determines scaling at a pseudo-critical point.
In a Griffiths-McCoy region frozen clusters produce already a finite
magnetization resulting in a classical low temperature behavior of the
susceptibility and specific heat. These override the confluent
singularities that are characterized by a continuously varying
exponent $z$ and are visible above a temperature $T^*\sim L^*\,\!^{-z}$.
\end{abstract}
\maketitle

The presence of quenched disorder in a quantum mechanical system may have
drastic effects in particular close to and at a quantum critical
point. The appearance of Griffiths-McCoy singularities
\cite{griffiths,mccoy}, leading to the divergence
of various quantities like the susceptibility at zero temperature even
far away from a quantum critical point, has received considerable
attention recently \cite{fisher,thill-huse,qsg-griff,igloi-rieger,pich}.
This quantum Griffiths behavior is characteristic for quantum phase
transitions described by an infinite randomness fixed point (IRFP)
\cite{fisher-irfp},which was shown to be relevant for many
disordered quantum systems \cite{monthus-igloi}.

Quantum Griffiths behavior was proposed to be the physical mechanism
responsible for the ``non-Fermi-liquid'' behavior observed in many
heavy-fermion materials \cite{jones,stewart}. However, it was also argued
that in a dissipative environment, as in metals due to the
conduction electrons, such a quantum Griffiths behavior might
essentially be non-existent \cite{castro,millis}. Moreover, even the
underlying sharp quantum phase transition itself was shown to be
rounded in dissipative model systems \cite{vojta}. Obviously there is
a need to examine carefully the effect of dissipation on a quantum
system displaying IRFP and quantum Griffiths behavior in the
non-dissipative case and to treat correctly the mixing of critical and
Griffiths-McCoy singularities --- which is what we intend to do in
this letter.

The properties of a single spin coupled to a dissipative bath has been
extensively examined \cite{spin_boson_review}. Upon increasing the
coupling strength between spin and bath degrees of freedom it displays
at zero temperature a transition from a non-localized phase, in which
the spin can still tunnel, to a localized phase, in which tunneling
ceases and the spin behaves classically.  Such a transition
is also present in an infinite ferromagnetic spin chain coupled to a
dissipative bath, as it was recently shown numerically \cite{troyer}.
Here we want to focus on the interplay of disorder, quantum
fluctuations and dissipation and study the Random Transverse Field
Ising Chain (RTIC) where each spin is coupled to an ohmic bath of
harmonic oscillators \cite{cugliandolo}. It is defined on a chain of
length $L$ with periodic boundary conditions (p.b.c.) and described by
the Hamiltonian $H$:
\begin{equation}
H=\sum^L_{i=1}
\biggl[-J_i \sigma_i^z \sigma_{i+1}^z - h_i \sigma_i^{x}
+ \sum_{k_i} \frac{p_{k_i}^2}{2} + 
\omega_{k_i} \frac{x_{k_i}^2}{2} + C_{k_i} x_{k_i} \sigma_i^z\biggr]
\label{Def_H} 
\end{equation}
where $\sigma_i^{x,z}$ are Pauli matrices and the masses of the
oscillators are set to one. The quenched random bonds $J_i$
(respectively random transverse field $h_i$) are uniformly distributed
between $0$ and $J_0$ (respectively between $0$ and $h_0$). The
properties of the bath are specified by its spectral function
$J_i(\omega) = \frac{\pi}{2} \sum_{k_i}
{C_{k_i}^2}/{\omega_{k_i}}\delta(\omega - \omega_{k_i}) =
\frac{\pi}{2} \alpha_i \omega e^{-\omega/\Omega_i}$
where $\Omega_i$ is a cutoff frequency. Initially the spin-bath
couplings and cut-off frequencies are site-independent, {\it i.e.}
$\alpha_i=\alpha$ and $\Omega_i=\Omega$, but both become
site-dependent under renormalization. 

To characterize the ground state properties of this
system~(\ref{Def_H}), we follow the idea of a real space
renormalization 
group (RG) procedure introduced in Ref. \cite{ma_dasgupta_rsrg} and pushed
further in the context of the RTIC without dissipation in
Ref. \cite{fisher}. The strategy is to find the largest coupling
in the chain, either a transverse field or a bond, compute the ground
state of the associated part of the Hamiltonian and treat the
remaining couplings in perturbation theory. The bath degrees of
freedom are dealt with in the spirit of the ``adiabatic
renormalization'' introduced in the context of the (single) spin-boson (SB)
model \cite{spin_boson_review}, where it describes
accurately its critical behavior~\cite{vojta_rg_num}.

Suppose that the largest coupling in the chain is a transverse field,
say $h_2$. Before we treat the coupling of site $2$ to the rest of
the system $-J_1\sigma_1^z\sigma_2^z-J_2\sigma_2^z\sigma_3^z$ 
perturbatively as in \cite{fisher}
we consider the effect of the part $-h_2\sigma_2^x+\sum_k
(p_k^2/2+\omega_k x_k^2/2 +C_k x_k\sigma_2^z)$ of the Hamiltonian,
which represents a single SB model. For this we integrate out
frequencies $\omega_k$ that are much larger than a lower cut-off
frequency $ph_2\ll\Omega_2$ with the dimensionless parameter
$p\gg1$. Since for those oscillators $\omega_k\gg h_2$ one can assume
that they adjust instantaneously to the current value of $\sigma_2^z$
the renormalized energy splitting is easily calculated within the
adiabatic approximation
\cite{spin_boson_review} and one gets an effective transverse field
$\tilde h_2<h_2$: 
\begin{equation}
\tilde h_2 = h_2 \left( {p h_2}/{\Omega_2} \right)^{\alpha_2}
\quad, \quad \tilde \Omega_2 = p h_2\;. \label{decim_h_1}
\end{equation}
If $\tilde h_2$ is still the largest
coupling in the chain the iteration (\ref{decim_h_1}) is repeated.
Two situations may occur depending
on the value of $\alpha_2$. If $\alpha_2 < 1$ this procedure
(\ref{decim_h_1}) will converge to a finite value 
$h_2^*=h_2(ph_2/\Omega_2)^{\alpha_2/(1-\alpha_2)}$
and the SB system at site $2$ is in a delocalized phase
in which the spin and the bath can be considered as being decoupled
(formally $\alpha_2=0$), as demonstrated by an RG treatment 
in \cite{vojta_rg_num}.

If this value $h_2^*$ is still the largest coupling in the
chain it will be aligned with the transverse field. As in the RTIC
without dissipation, this spin is then decimated (as it will not
contribute to the magnetic susceptibility) and gives rise, in second
order degenerate perturbation theory, to an effective coupling
$\tilde{J_1}$ between the neighboring moments at site $1$ and
$3$~\cite{fisher}
\begin{eqnarray}
\tilde J_1 = J_1 J_2/h_2^* \label{decim_h_2}
\end{eqnarray}
If $\alpha_2 >1$, $\tilde h_2$ can be made arbitrarily small by
repeating the procedure (\ref{decim_h_1}) implying that the 
SB system on site $2$ is in its localized phase 
\cite{vojta_rg_num} and essentially behaves classically:
the decimation rule (\ref{decim_h_1}) indeed amounts to set
$\tilde{h}_2=0$. Such a moment, or cluster of spins, will be aligned
with an infinitesimal external longitudinal field and is
denoted as ``frozen''.

Suppose now that the largest coupling in the chain is a bond, say
$J_2$. The part of the Hamiltonian that we focus on is
$-J_2\sigma_2^z\sigma_3^z+\sum_{i=2,3}\sum_{k_i} (p_{k_i}^2/2+\omega_{k_i}
x_{k_i}^2/2 + C_{k_i} x_{k_i}\sigma_i^z)$, {\it i.e} a subsystem of
two spin-bosons coupled via $J_2$. We find that in second order
perturbation theory the ground state of this subsystem is equivalent
to a single SB system coupled to {\it both} baths leading to
the additive rule
\begin{eqnarray}
\tilde \alpha_2 = \alpha_2+\alpha_3 \label{add_alpha}
\end{eqnarray} 
Integrating out the degrees of freedom of both baths with frequencies
larger than $p J_2$ (as done previously for the single SB system
\cite{vojta_rg_num}) the two moments at $2$ and $3$
are replaced by a single moment $\tilde{\mu}_2$ with an effective
transverse field $\tilde{h}_2$:
\begin{eqnarray}
&&\tilde h_2 = \frac{h_2 h_3}{J_2} \left(\frac{p J_2}{{\Omega}_2}
\right)^{\alpha_2}  \left(\frac{p J_2}{{\Omega}_3}
\right)^{\alpha_3} \label{decim_J_1} \\ 
&&\tilde \mu_2 = \mu_2+\mu_3 \quad, \quad   \tilde \Omega_2 = p J_2
\label{decim_J_2} 
\end{eqnarray}
where $\mu_i$ is the magnetic moment of site $i$ (in the original
model, one has $\mu_i=1$ independently of $i$). Combining
Eq.~(\ref{add_alpha}) and Eq.~(\ref{decim_J_2}) one clearly sees that
$\tilde \alpha_i = \tilde \mu_i \alpha$. 

In the following we analyze this RG procedure defined by the
decimation rules (\ref{decim_h_1}-\ref{decim_J_2}) numerically. This
is done by considering a finite chain of size $L$ with p.b.c. and
iterating the decimation rules until only one site is left. This
numerical implementation has been widely used in previous works
\cite{monthus-igloi}, and it has been shown in particular to reproduce
with good accuracy the exact results of Ref. \cite{fisher} for the
RTIC. We fix $h_0 = 1$ and concentrate on the parameter plane
$(\alpha, J_0)$. All data were obtained by averaging over $10^5$
different disorder realizations (if not mentioned otherwise), and the
disorder average of an observable $\cal{O}$ is denoted by
$\overline{\cal{O}}$. The decimation rules
(\ref{decim_h_1}-\ref{decim_J_2}) depend explicitly on the ``ad hoc''
parameter $p$ (or more precisely on the ratio $\Omega/p$). For the
moment we fix $\Omega/p = 10^4$ and discuss the weak dependence on
this parameter below.

\begin{figure}
\includegraphics[width=\linewidth]{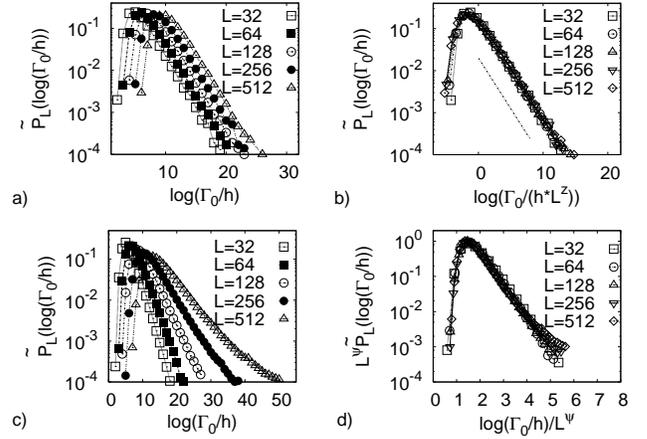}
\caption{{\bf a)} $\tilde P_L(\log(\Gamma_0/h))$ as a function of
  $\log(\Gamma_0/h)$ for different system sizes $L$ for
  $\alpha=0.03$ ($h_0=1$, $J_0=0.34$), {\it i.e.} far
  from the pseudo-critical point. 
 {\bf b)} $\tilde P_L(\log(\Gamma_0/h))$ as a function of $\log(\Gamma_0/h
  L^z)$ for different system sizes for $\alpha= 0.03$ ($h_0=1$,
  $J_0=0.34$). The straight dashed line has slope $1/z$ with $z=1.65(5)$. 
{\bf c)} $\tilde P_L(\log(\Gamma_0/h))$ as a function of
  $\log(\Gamma_0/h)$ for different system sizes $L$ for
  $\alpha=0.052$ ($h_0=1$, $J_0=0.34$), {\it i.e.} at the
  pseudo-critical point. {\bf d)} $L^\psi \tilde
  P_L(\log(\Gamma_0/h))$ as a function of 
  $\log(\Gamma_0/h)/L^\psi$ for different system sizes for $\alpha =
  0.052$, $\psi=0.32$.
}
\label{fig_griffith}  
\end{figure}

The transverse field $h$ acting on the last remaining spin is an
estimate for the smallest excitation energy. Its distribution, 
$P_L(h/\Gamma_0)$, where $\Gamma_0$ is the largest coupling in
the initial chain of size $L$, reflects the characteristics of the gap
distribution \cite{igloi-rieger}. Since the last spin can either be frozen 
(i.e the last field $h$ is zero) or non-frozen we split 
$P_L(h/\Gamma_0)$ into two parts:
\begin{eqnarray}
\tilde P_L(h/\Gamma_0) = A_L \tilde P_L(h/\Gamma_0) + (1-A_L)
\delta(h/\Gamma_0) \label{gen_form} 
\end{eqnarray}
where $\tilde P_L(h/\Gamma_0)$ is the restricted distribution of the
last fields in the samples that are non-frozen and $A_L$ is the
fraction of these samples. It, or equivalently $\tilde
P_L(\log(\Gamma_0/h))$, represents the distribution of the
smallest excitation energy in the ensemble of non-localized spins.

\begin{figure}
\includegraphics[width=\linewidth]{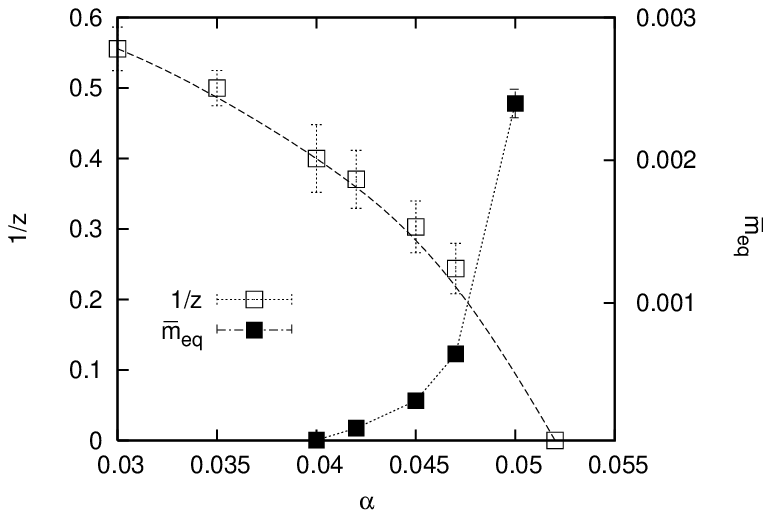}
\caption{Magnetization $\overline m_{\text{eq}}$ and inverse dynamical exponent
  $1/z$ as a function of $\alpha$ (for $h_0=1$, $J_0=0.34$)
} \label{comb_z_m}
\end{figure}

Let us first present data obtained for $J_0 = 0.34$. 
Fig. \ref{fig_griffith}a shows $\tilde
P_L(\log(\Gamma_0/h))$, for $\alpha=0.03$. For a system close to, but
not at, a quantum critical point described by an IRFP one expects
indications of Griffiths-McCoy singularities characterized by the
following scaling behavior for $\tilde P_L$
\begin{eqnarray}
\tilde P_L(\log(\Gamma_0/h)) = {\cal P}(\log(\Gamma_0/h L^z))\;, 
\label{griffith_scaling}
\end{eqnarray}
where $z$ is a dynamical exponent continuously varying with ($J_0$,
$\alpha$, etc.). Fig. \ref{fig_griffith}b shows a good data collapse
with $z=1.65$ for the chosen coupling constant $\alpha=0.03$. The
slope of the dotted line in Fig. \ref{fig_griffith}b is identical to
$1/z$ and upon increasing $\alpha$ we observe that the slope, $1/z$, 
decreases. Our numerical estimates for $1/z(\alpha)$ are
shown in Fig. \ref{comb_z_m}; they indicate that $1/z$ approaches zero
for some critical value $\alpha_c$, which implies formally $z\to
\infty$ for $\alpha\to\alpha_c$. This is also characteristic for an 
IRFP, where $\tilde P_L(\log(\Gamma_0/h))$ is expected to scale as
\begin{eqnarray}
\tilde P_L(\log(\Gamma_0/h)) = 
L^{-\psi}{\cal P}_{\text{IRFP}}(L^{-\psi}\log(\Gamma_0/h)\;.
\label{irfp_scaling} 
\end{eqnarray}
$\psi$ is a critical exponent characterizing the IRFP.
Fig. \ref{fig_griffith}c shows $\tilde P_L(\log(\Gamma_0/h))$ for
$\alpha=0.052$: 
one observes that it broadens systematically with
increasing system size, in contrast to the data shown in
Fig. \ref{fig_griffith}a, and Fig. \ref{fig_griffith}d displays a good data
collapse according to (\ref{irfp_scaling}) with $\psi=0.32$. Varying
$\alpha$ only slightly worsens the data collapse substantially,
therefore we take $\alpha_c=0.052$ as our estimate for the putative 
critical point (for $h_0=1$ and $J_0=0.34$) and denote by $\Delta =
(\alpha_c-\alpha)/\alpha_c$ the distance from it.

\begin{figure}
\includegraphics[width=\linewidth]{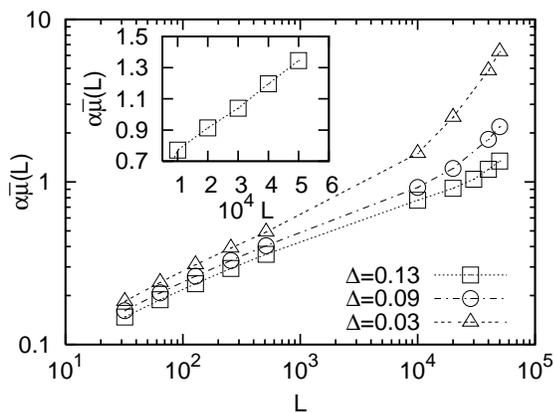}
\caption{Magnetic moment $\alpha \overline{\mu}(L)$ as a function of
  the system size $L$ for different values of $\delta$. {\bf Inset} :
  $\alpha \overline{\mu}(L)$ for $L \geq 10^4$ as a function of
  $L$ and for $\Delta = 0.13$. The linear behavior suggests a finite
  $\overline m_{eq}$. The lines are guides to the eyes.
  Due to the large system sizes the data are averaged only over
  $500$ different disorder realizations.) 
}\label{fig_mu}
\end{figure}

The magnetic moment $\mu$ of the last remaining spin in
the decimation procedure represents an estimate of the total
magnetization $\overline m_{\text{eq}}L$ of the chain. 
In Fig. \ref{fig_mu}, we show $\alpha \overline \mu(L)$ as a function
of $L$ for different values of $\Delta$. For small $L$, $\overline \mu(L)
\propto L^a$ with $a \simeq 1/3$ up to a length scale $L^*\sim{\cal O}(10^4)$
beyond which the effective coupling between strongly coupled clusters
and the bath, $\alpha \overline \mu$, gets larger than one and the clusters
become localized. Above this value $\overline \mu(L) \sim \overline
m_{\text{eq}} L$ (see inset of Fig. \ref{fig_mu}), which suggests a
finite magnetization $\overline m_{\text{eq}}$ {\it before} the
putative critical point is reached. This is a manifestation of the
``frozen'' clusters and lead to the concept of rounded quantum phase
transitions in the presence of dissipation \cite{vojta}. The typical
size of a frozen cluster turns out to be rather large $L^*\geq10^4$ for
this range of parameters $(\alpha=0.03-0.052,J_0=0.34)$. Consequently
the fraction of non-frozen samples, $A_L$, in (\ref{gen_form}) is
close to~$1$ for the system sizes that we could study numerically.

\begin{figure}
\includegraphics[width=\linewidth]{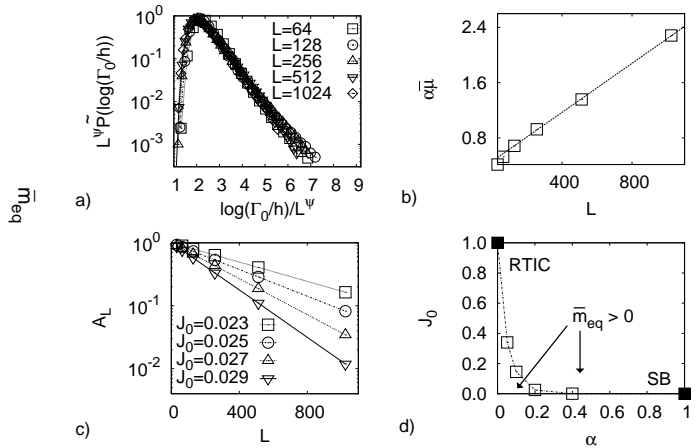}
\caption{{\bf a)} $L^\psi \tilde P_L(\log(\Gamma_0/h))$ as a function of
  $\log(\Gamma_0/h)/L^\psi$ with $\psi=0.31$ for different system
  sizes for $\alpha = 
  0.2$ and $J_0 = 0.025$. {\bf b)} Magnetic moment $\alpha
  \overline{\mu}(L)$ as a function of 
  the system size $L$ for $\alpha=0.2$ and $J_0 = 0.025$ suggesting
  $\overline m_{\rm eq} > 0$ and $L^*\sim  
  100$. {\bf c)}
  $A_L$ as a function of $L$ on a linear-log plot for different values
  of $J_0$ and $\alpha=0.2$. {\bf d)} Phase diagram for $h_0=1$ and
  $\Omega/p = 10^4$ characterized by a single phase with $\overline m_{\rm
    eq}>0$. The dotted line represents the line of smeared transitions
  characterized by an IRFP scaling (\ref{gen_form},\ref{irfp_scaling}).
}\label{combined_fig}  
\end{figure}

A stronger dissipation strength $\alpha$ reduces $L^*$ and gives us
the possibility to study the crossover to a regime that is dominated
by frozen clusters, in particular the $L$-dependence of $A_L$ in
(\ref{gen_form}), and we consider $\alpha=0.2$ as an example now. For
the restricted distribution $\tilde P(\log(\Gamma_0/h))$, we obtain
the same scenario as for smaller dissipation, as shown in
Fig. \ref{combined_fig}a for the putative critical point $J_{0c} =
0.025$. Fig. \ref{combined_fig}b , shows $\alpha \overline
\mu(L)$ indicating that $\overline \mu(L)\propto \overline
m_{\text{eq}} L$ for $L>100$, which implies $L^*\sim 100$.  The
fraction of non-frozen samples, $A_L$, shows a clear deviation from
unity already for the system sizes we study here:
Fig. \ref{combined_fig}c shows $A_L$ as a function of $L$ for
different values of $J_0$. The data imply an exponential decay $A_L
\propto e^{-L/\tilde L}$. The characteristic decay length fits well to
$\tilde L\propto J_0^{\alpha^{-\beta}}$, with $\beta \simeq 0.8$, meaning a
very rapid increase of $\tilde{L}$ with decreasing dissipation
strength $\alpha$. By comparing $\tilde{L}$ with $L^*$ for various
parameters $(\alpha, J_0)$ we find that $\tilde L = \kappa L^*$, with
$\kappa$ a dimensionless number of order one, weakly dependent on
$\alpha$ and $J_0$. 

As long as $L<L^*$ the restricted distribution is not significantly
different from the full distribution of non-vanishing excitation
energies, since the probability for a frozen sample is small for $L\ll
L^*$. Since $\tilde{P}_L$ has a power law tail down to excitation
energies exponentially small in $L$, the specific heat, susceptibility
etc.\ in finite size systems display a singular low temperature
behavior characterized by the dynamical exponent $z(\alpha)$ down to
very low temperatures (actually down to $T_L\sim e^{-aL}$). This
intermittent singular behavior, $\chi(T)\sim T^{-1+1/z(\alpha)}$ for
the susceptibility and $c(T)\sim T^{1/z(\alpha)}$ for the specific
heat, persists for larger system sizes as well as for $L\to\infty$,
but as soon as $L>L^*$ it will eventually compete with the temperature
dependence of the (quantum mechanically) frozen clusters - e.g. $1/T$
for the susceptibility. Since the latter has a small amplitude
proportional to $1/L^*$, classical temperature dependence will only
set in below $T^*\sim L^*\,\!^{-z(\alpha)}$ and Griffiths-like
behavior is visible (also in the infinite system) above $T^*$.

It is instructive to consider the RTIC without dissipation, but with a
finite fraction $\rho$ of zero transverse fields (i.e.\
$p(h)=\rho\delta(h)+h_0^{-1}(1-\rho)\theta(h)\theta(h_0-h)$). The
sites with $h=0$ then correspond to frozen
clusters that have an average distance $L^*\propto\rho^{-1}$. Indeed
the distribution $P(h/\Gamma_0)$ shows the same behavior as in
Eq.~(\ref{gen_form}) with $A_L \sim e^{-L/L^*}$. But, in contrast to
the dissipative case, the restricted distribution $\tilde
P(\log{(\Gamma_0/h)})$ is naturally identical with the one for the
non-diluted ($\rho=0$) RTIC, which shows the IRFP scaling
(\ref{irfp_scaling}) at $h_0=J_0$ with $\psi_{\rm{RTIC}} = 0.5$,
different from the one we obtain here.

The connected correlation function $\overline C(r)= 
\overline{ \langle\sigma_i^z\sigma_{i+r}^z\rangle - \langle
\sigma_i^z \rangle \langle \sigma_{i+r}^z \rangle }$ 
decays exponentially for $r \gg L^*$, given that the quantum
fluctuations are exponentially suppressed beyond this length scale
(\ref{gen_form}), consistent with \cite{vojta}. It should also be
noted that the connected correlation function of the restricted
ensemble of non-frozen samples $\overline{\tilde C}(r)$ does not
behave critically since the number of non frozen spins belonging to
the same cluster is bounded by $1/\alpha$ in the restricted
ensemble. Thus, the origin of the systematic broadening of the
distribution $\tilde P(\log{\Gamma_0/h})$ is here different from a
standard IRFP and probably stems from the non-localized clusters with
$\alpha_i$ close to (but smaller than) one (see Eq. \ref{decim_h_1}).

We have checked that the behavior of the gap distribution
characterized by Eq. (\ref{gen_form},\ref{irfp_scaling}) depends very
weakly on the {\it ad hoc} parameter $\Omega/p$ in the range
$10-10^4$. In this range, the relative variations of the estimated
exponent $\psi$ is of the order of $5 \%$, although the values of
$L^*$ and $(\alpha_c, J_{0c})$ are more sensitive, and probably non
universal. We repeated the previous analysis for different values of
$(\alpha, J_0)$ (keeping $h_0=1$). In contrast to the pure case
\cite{troyer}, the entire plane is here characterized by a single phase 
where $\overline{m}_{\text{eq}} > 0$, beyond a length scale $L^*\equiv
L^*(\alpha,J_0)$, everywhere (except on the boundaries $(\alpha,0)$
and $(0,J_0)$) \cite{footnote_1}. One can identify a
line of smeared transitions associated with the broadening of the
restricted gap distribution $\tilde P(\log{(\Gamma_0/h)})$, according
to (\ref{irfp_scaling}): this is depicted in Fig. \ref{combined_fig}d. 
We find that the associated exponent $\psi$ vary weakly along this
line, its relative variation being less than $10\%$.

To conclude our strong disorder RG study of the RTIC coupled to a
dissipative Ohmic bath revealed that non-frozen samples display an
IRFP scaling of the distribution of excitation energies. With this we
computed a continuously varying exponent $z(\alpha)$ that determines
an intermittent singular temperature dependence of thermodynamic
quantities above a temperature $T^*\sim L^*\,\!^{-z(\alpha)}$. $L^*$
is a characteristic length scale above which the ground state displays
a non-vanishing magnetization, as predicted by the smeared transition
scenario \cite{vojta}, and that we determined to increase
exponentially with the inverse strength of the dissipative
coupling. This implies that numerical studies can hardly track the
asymptotic behavior \cite{cugliandolo} and that experiments at very
low but non-vanishing temperatures might still show indications for
quantum Griffiths behavior \cite{jones,castro,millis}. In higher
dimensions we expect a similar scenario as the one discussed here
and it would be interesting to extend our study to Heisenberg
antiferromagnets and XY systems.

\acknowledgments
HR thanks the Aspen Center for Physics, where parts of this work were
done, for its kind hospitality, and T. Vojta for stimulating
discussions. GS acknowledges the financial support provided through
the European Community's Human Potential Program under contract
HPRN-CT-2002-00307, DYGLAGEMEM.


\begin{thebibliography}{35}
\bibitem{footnote_1}
Computing the dependence of $\overline m_{\rm eq}$ as a function
of the distance from the pure critical point \cite{vojta} requires a precise
estimate of the critical line itsel, which is not feasible with our method.
\end{thebibliography}
\end{document}